\documentstyle[preprint,tighten,aps,graphicx,floats]{revtex}
\def\ba{\begin{eqnarray}}
\def\ea{\end{eqnarray}}
\def\bq{\begin{equation}}
\def\eq{\end{equation}}
\def\lsim{\mathrel{\raisebox{-.6ex}{$\stackrel{\textstyle<}{\sim}$}}}
\def\sla#1{\ifmmode%
\setbox0=\hbox{$#1$}%
\setbox1=\hbox to\wd0{\hss$/$\hss}\else%
\setbox0=\hbox{#1}%
\setbox1=\hbox to\wd0{\hss/\hss}\fi%
#1\hskip-\wd0\box1 }

\begin{document}
\thispagestyle{empty}

\renewcommand{\small}{\normalsize} 

\preprint{
\font\fortssbx=cmssbx10 scaled \magstep2
\hbox to \hsize{
\hskip.5in \raise.1in\hbox{\fortssbx University of Wisconsin - Madison}
\hfill\vtop{\hbox{\bf MADPH-00-1208}
            \hbox{\bf FERMILAB-Pub-00/332-T}
            \hbox{hep-ph/0012351}
            \hbox{December 2000}} }
}

\title{\vspace{.45in}
$H\to WW$ as the discovery mode for a light Higgs boson
}
\author{N.~Kauer$^1$, T.~Plehn$^1$, D.~Rainwater$^2$ and 
D.~Zeppenfeld$^1$\\[3mm]}
\address{
$^1$Department of Physics, University of Wisconsin, Madison, WI 53706\\
$^2$Fermi National Accelerator Laboratory, Batavia, IL 60510
}
\maketitle
\begin{abstract}
The production cross section for a $m_H\approx 115$~GeV, SM Higgs boson 
in weak boson fusion at the LHC is sizable. However, the branching fraction
for $H\to W^+W^-$ is expected to be relatively small. 
The signal, with its two forward jets, is sufficiently different from
the main backgrounds that a signal to background ratio of better than 1:1
can nevertheless be obtained, with large enough rate to allow for a $5\sigma$
signal with $35$~fb$^{-1}$ of data. The $H\to WW$ signal in weak boson fusion
may thus prove to be the discovery mode for the Higgs boson at the LHC.
\end{abstract}

\vspace{0.4in}


After the end of LEP~II running, the search for the Higgs boson and, hence, 
for the origin of electroweak 
symmetry breaking and fermion mass generation, remains one of the premier 
tasks of present and future high energy physics experiments. With the
exclusion of a SM Higgs boson of mass $m_H<113.5$~GeV and some evidence
for $m_H\approx 115$~GeV\cite{lep2higgs}, the mass range 
now preferred by supersymmetry\cite{one_loop,two_loop}, 113~GeV$\lsim m_h 
\lsim 130$~GeV, will likely remain the focus of investigations until the
beginning of data taking at the CERN LHC.

Previously we have shown that a somewhat heavier Higgs boson, in the range
130~GeV$\lsim m_H \lsim 190$~GeV, will give a highly significant $H\to W^+W^-
\to e^\pm\mu^\mp\sla p_T$ signal in weak boson fusion (WBF) at the 
LHC\cite{RZ_WW,thesis}. For a smaller Higgs boson mass, the branching ratio
of the $H\to W^+W^-$ mode quickly decreases, making the observation of this
mode more difficult. Only a marginal signal was expected 
for $m_H\approx115$~GeV. However, this earlier analysis 
was optimized for a Higgs boson mass near $W$-pair threshold and can clearly
be improved for the significantly smaller masses favored at the end of LEP~II
running. This situation motivates a complete reanalysis of possible
$H\to W^+W^-$ signals in WBF at the LHC, and of possible backgrounds, 
with the goal of optimizing the significance of a signal for 
$m_H\approx 115$~GeV. At the same time a better signal to background $(S/B)$
ratio, higher signal acceptance and higher accuracy in the determination
of $B\sigma(qq\to qqH,\;H\to WW)$ would significantly improve the extraction
of Higgs boson properties such as its total width or the $HWW$ 
coupling constant\cite{zknr}.

In this letter we summarize the results of this reanalysis, and show that
even for a SM Higgs of $m_H=115$~GeV the $H\to WW$ decay mode can be observed
in WBF, with better than $5\sigma$ significance, with about 35~fb$^{-1}$ of 
LHC data. 
This result of our parton level analysis, if confirmed by a more complete
simulation of hadronization and detector effects, would render this 
WBF mode the most promising single search channel at the LHC, superior even
to the classic inclusive search for a $H\to\gamma\gamma$ resonance 
peak~\cite{CMS,ATLAS}. 

Our results are based on parton level simulations of the signal and the 
various backgrounds, with full tree level matrix elements of all contributing
subprocesses. Thus, final state partons are identified with observable jets.
The signal can be described as the scattering of quarks and/or
anti-quarks via $t$-channel $W$ and $Z$-exchange, with the Higgs boson 
radiated off this weak boson. The signal contains two (forward) quark-jets,
called tagging jets, in addition to the $H\to W^+W^-\to ll'\nu\bar\nu$
decay products. Here, one or both $W$ bosons may be virtual. 
We consider $ll'=e^\pm\mu^\mp$ and also 
$ll'=e^+e^-,\;\mu^+\mu^-$ final states. The latter, dubbed 'same flavor'
or $ll$ sample, were not considered in Refs.~\cite{RZ_WW,thesis}.

Any processes resulting in two jets, two (oppositely) charged leptons and
missing transverse momentum constitute potential backgrounds. The dominant
background turns out to be top-quark pair production in association with
up to two additional light-quark or gluon jets. For the $t\bar t$ and 
$t\bar tj$ simulation we use the results of Ref.~\cite{kauer} which include 
off-shell top and $W$ effects and take into account the single-resonant and 
non-resonant contributions. This inclusion is potentially
important since, for the Higgs masses considered here, the $ll'\nu\bar\nu$
final state is produced well below threshold for $t\bar t$ decays.
The top-quark backgrounds are separated into three categories, depending on 
whether two, one or zero $b$ (or $\bar b$) quarks are identified as one of
the two forward
tagging jets, and are called $tt$, $ttj$ and $ttjj$ backgrounds, respectively.

$W$-pair production in association with two light partons constitutes the
next major background. The 'QCD $WWjj$' background is calculated at order
${\cal O}(\alpha^2\alpha_s^2)$ and contains the real emission corrections 
to $q\bar q\to W^+W^-$
(and crossing related processes). The 'EW $WWjj$' background consists of
all $q_1q_2\to q_3q_4W^+W^-$ processes at order ${\cal O}(\alpha^4)$.
Similarly, the 'QCD $\tau\tau jj$' and 'EW $\tau\tau jj$' backgrounds
capture $\tau$-pair production in association with two jets at 
${\cal O}(\alpha_s^2\alpha^2)$ and ${\cal O}(\alpha^4)$, respectively,
with subsequent leptonic $\tau$-decay. They include full $Z,\gamma$ 
interference effects. Finally, the '$bbjj$' background arises from 
$b\bar bjj$ production (simulated at ${\cal O}(\alpha_s^4)$), with 
semi-leptonic decays of both bottom quarks. The simulation of these 
backgrounds is performed as in Refs.~\cite{RZ_WW,RZ_tautau_ll}.
For the same flavor sample we also consider QCD $ZZjj$ final states with
$ZZ\to l^+l^- \nu\bar\nu$ (generated like the QCD $WWjj$ background and 
including $\gamma^*\to l^+l^-$ contributions); and QCD
and EW $lljj$ final states, which are calculated like the $\tau\tau jj$ 
backgrounds, except that the missing transverse momentum is now entirely
due to detector effects, which are simulated as in Ref.~\cite{RZ_WW}.

The basic event selection which we propose is similar to the one suggested 
previously~\cite{RZ_WW}. One looks for events with at least two jets 
(tagging jets) and two 
charged leptons in the phase space region
\ba
\label{eq:basic}
& p_{Tj} \geq 20~{\rm GeV} \, , \; \; |\eta_j| \leq 4.5 \, , \; \;
\triangle R_{jj} \geq 0.6 \, , \nonumber\\
p_{T\ell_1} \geq 20~{\rm GeV} \, ,\; & p_{T\ell_2} \geq 10~{\rm GeV}\;,\quad
|\eta_{\ell}| \leq 2.5 \, , \; \; \triangle R_{j\ell} \geq 1.7 \, .
\ea
The staggered lepton $p_T$ cut accommodates the softer lepton from the 
decay of a virtual $W$. Both charged leptons must lie between the tagging 
jets with a separation in pseudorapidity $\triangle \eta_{j,\ell} > 0.6$, 
and the jets must occupy opposite hemispheres:
\ba
\label{eq:lepcen}
& \eta_{j,min} + 0.6 < \eta_{\ell_{1,2}} < \eta_{j,max} - 0.6 \, ,
\nonumber \\
& \eta_{j_1} \cdot \eta_{j_2} < 0 \;.
\ea
A large dijet invariant mass and a wide separation in pseudorapidity is 
required for the two forward tagging jets,
\begin{equation}
\label{eq:gap}
m_{jj}>600~{\rm GeV}\, ,\qquad 
\triangle \eta_{tags} = |\eta_{j_1}-\eta_{j_2}| \geq 4.2 \, ,
\end{equation}
leaving a gap of at least 3 units of pseudorapidity in which the charged 
leptons can be observed. 

Forward jet tagging has been discussed as an effective technique to separate 
weak boson scattering from various  backgrounds in the past~
\cite{Cahn,BCHP,DGOV,bjgap,RSZ_vnj,RZ_WW,thesis},
in particular for heavy Higgs boson searches. A second technique for 
suppression of QCD backgrounds to WBF is the veto of any additional 
identifiable jet activity in the central region\cite{bjgap,RSZ_vnj}. 
We discard all events where an additional veto jet of $p_{Tv} > 20$~GeV is 
located in the gap region between the two tagging jets,
\begin{equation}
\label{eq:bveto}
p_{Tv} > 20\;{\rm GeV} \, , \; \;
\eta_{j,min} < \eta_{v} < \eta_{j,max} \, .
\end{equation}
First of all the jet veto is very effective against the $t\bar t$ backgrounds,
by vetoing most of the relatively hard $b$ and $\bar b$ arising from top-quark
decay~\cite{CMS,ATLAS}. 
In addition, the central jet veto exploits the different gluon 
radiation patterns of WBF, where most additional partons will be emitted in the
far forward and far backward directions, and of $t$-channel gluon exchange
processes, which prefer additional parton emission in the central region.
We do not explicitly generate these additional soft jets in our parton level
Monte Carlo programs. Their effect has been estimated in 
Refs.~\cite{RZ_WW,thesis,RSZ_vnj} and can be included as overall veto survival
probabilities of $P_{surv}=0.89$ for the Higgs signal, 0.29 for the QCD
backgrounds (including $t\bar t$ production) and 0.75 for the EW backgrounds. 
There is some uncertainty in these estimates and they will eventually have to
be determined experimentally, from LHC data on $Wjj$ and $Zjj$ 
production~\cite{RSZ_vnj}. We therefore include these survival probabilities
only at the final stage of the analysis.

The two (virtual) $W$s of the Higgs boson signal are produced close to
threshold and are almost at rest in the Higgs frame. As a result,
the charged lepton and the neutrino arising from a single $W$ are almost
back-to-back  in this frame, and they have equal energies. This implies
that the invariant masses of the two charged leptons and of the two neutrinos
are approximately equal, $m_{ll}\approx m_{\nu\bar\nu}$, and neither can exceed
half the Higgs boson mass. In the lab frame the relatively small dilepton 
invariant mass favors a small angle between the two charged leptons. Since
we are interested in Higgs boson masses $m_H\lsim 130$~GeV, we require
\begin{equation}
\label{eq:mll.phi}
m_{ll} < 60\;{\rm GeV} \, , \; \;
\phi_{ll} < 140^o \, ,
\end{equation}
where $\phi_{ll}$ is the azimuthal angle between the charged leptons. Small 
angular separations of the leptons are further favored by the tensor
structure of the SM $HWW$ coupling~\cite{DittDrein,RZ_WW}. These cuts still
leave a large background from $\tau$-pairs. For $\tau^+\tau^-$ events the
energy fractions $x_{\tau_1},\;x_{\tau_2}$ carried by the observed 
$\tau$-decay leptons can be determined from transverse momentum balance,
${\bf p}_{T\tau_1}+{\bf p}_{T\tau_2}={\bf p}_{Tl_1}+{\bf p}_{Tl_2}+
{\sla{\bf p}}_T$, and thus the $\tau$-pair invariant mass, $m_{\tau\tau}=
m_{ll}/\sqrt{x_{\tau_1}x_{\tau_2}}$ can be reconstructed~\cite{tautaumass}. 
We veto any events consistent with $Z\to\tau\tau$, i.e. events which 
satisfy the three conditions
\begin{equation}
\label{eq:xtau.mtau}
x_{\tau_1}>0\;,\qquad x_{\tau_2}> 0\;, \qquad 
m_{\tau\tau} > m_Z -25\;{\rm GeV} \, .
\end{equation}

Positive identification of the Higgs boson is greatly aided by the 
reconstruction of a Higgs mass peak. Because of the two missing neutrinos
in the decay of the two $W$s the invariant mass of the Higgs decay products
cannot be reconstructed.\footnote{In the threshold approximation, i.e. 
neglecting the $W^\pm$ momenta in the Higgs rest frame, and assuming one
$W$ to be on-shell, the Higgs mass can be reconstructed, with, typically,
a two-fold ambiguity. We find that this reconstruction is less efficient for 
rejecting backgrounds than the $WW$ transverse mass.}
However, it is possible to reconstruct the transverse
mass of the $ll\sla p_T$ system by using the threshold relationship 
$m_{\nu\nu}=m_{ll}$.  The transverse mass can then be defined as~\cite{RZ_WW}
\bq
M_T(WW) = 
\sqrt{({\sla{E}_T}+E_{T,ll})^2 - ({\bf p}_{T,ll}+{\sla{\bf p}}_T)^2}
\, ,
\label{eq:M_T.def}
\eq
with the transverse energies given by
$E_{T,ll} = \sqrt{{\bf p}_{T,ll}^2 + m_{ll}^2}$ and 
$\sla{E}_T = \sqrt{{\sla{\bf p}}_T^2 + m_{ll}^2}$. The Higgs 
signal is largely concentrated in the region
\bq
\label{eq:MTWW.cut}
50~{\rm GeV} < M_T(WW) < m_H+20~{\rm GeV}\; .
\eq

The first column of Table~I gives the cross sections for a Higgs signal of 
mass 115~GeV and the various backgrounds within the cuts of 
Eqs.~(\ref{eq:basic}--\ref{eq:MTWW.cut}). In addition to a sizable
$t\bar t+$jets background, the $bbjj$ background sticks out. It arises
from semi-leptonic $b\to cl\nu$ decays in which little energy is carried 
by the charm quark and other hadronization products. There are many ways
to reduce this background, e.g. by imposing harder cuts on $\sla p_T$, or the
transverse momenta of the tagging jets or the reconstructed Higgs, by looking
for displaced decay vertices, by tightening lepton isolation cuts, etc.

\begin{table}[t]
\caption{Signal and background cross sections (in fb) within the cuts
indicated. The first four columns refer to $e\mu\sla p_T$ final states,
while the last column gives results, after all the cuts of 
Eqs.~(\ref{eq:basic}--\ref{eq:cuts.ll}), for the $ll=ee,\;\mu\mu$ sample. 
The survival probabilities for a central jet veto are factored into the 
cross sections of the last two columns. In the first three columns the central
jet veto is only applied to $b$-quark jets from top decay.}
\label{tab:channels}
\begin{tabular}{l||c|c|c|c||c}
 channel &
 Eqs.~(\ref{eq:basic}--\ref{eq:MTWW.cut}) & 
 $+$~Eq.~(\ref{eq:contour}) &
 $+$~Eq.~(\ref{eq:emucut})  & 
 $\times P_{surv}$ &
 same flavor \\ \hline 
 $m_H=115$~GeV            & 1.04  & 0.93  & 0.92  & 0.83   & 0.58    \\ 
 $t\bar{t}$               & 0.051 & 0.040 & 0.040 & 0.012  & 0.010   \\ 
 $t\bar{t}j$              & 1.54  & 1.34  & 1.33  & 0.39   & 0.29    \\ 
 $t\bar{t}jj$             & 0.31  & 0.27  & 0.27  & 0.078  & 0.065   \\ 
 $b\bar{b}jj$             & 20.8  & 0.48  & 0.013 & 0.004  &$<0.001$ \\ 
 QCD $WWjj$               & 0.26  & 0.23  & 0.23  & 0.066  & 0.054   \\ 
 EW  $WWjj$               & 0.19  & 0.17  & 0.17  & 0.125  & 0.103   \\ 
 QCD $\tau\tau jj$        & 0.67  & 0.11  & 0.10  & 0.032  & 0.028   \\ 
 EW  $\tau\tau jj$        & 0.118 & 0.026 & 0.024 & 0.018  & 0.018   \\ \hline
 QCD $\ell\ell jj$        &       &       &       &        & 0.086   \\ 
 EW  $\ell\ell jj$        &       &       &       &        & 0.016   \\
 $ZZjj$                   &       &       &       &        &$<0.001$ \\
\end{tabular}
\end{table}

Within our parton level simulation we find that the most efficient way to
effectively eliminate the $bbjj$ and also the $\tau\tau jj$ backgrounds is
to exploit correlations between lepton azimuthal angles and the transverse
momentum of the reconstructed Higgs boson, 
$p_{TH} = |{\bf p}_{Tl_1}+{\bf p}_{Tl_2}+{\sla{\bf p}}_T|$. In $b\bar b$ and 
$\tau^+\tau^-$ decays, missing transverse momentum arises
from neutrinos which are emitted parallel to the observed 
charged leptons. As a result the ${\sla{\bf p}}_T$-vector lies between the
two lepton transverse momentum vectors, and, hence, close to their sum,
${\bf p}_{Tll} = {\bf p}_{Tl_1}+{\bf p}_{Tl_2}$. In the Higgs signal, on 
the other hand, the two leptons are emitted close to each other in the 
Higgs rest frame~\cite{DittDrein} with the neutrinos recoiling against
them. These features are captured by the azimuthal angle, 
$\Delta\phi(ll,\sla p_T)$, between ${\bf p}_{Tll}$ and ${\sla{\bf p}}_T$.
The $bbjj$ and $\tau\tau jj$ backgrounds are concentrated at small values of
$\Delta\phi(ll,\sla p_T)$ while the signal favors large separations, except 
when a large transverse momentum boost of the Higgs decay products moves 
them close together. 

\begin{figure}[tb]
\vspace*{-2.5cm}
\begin{center}
\includegraphics[width=14.0cm]{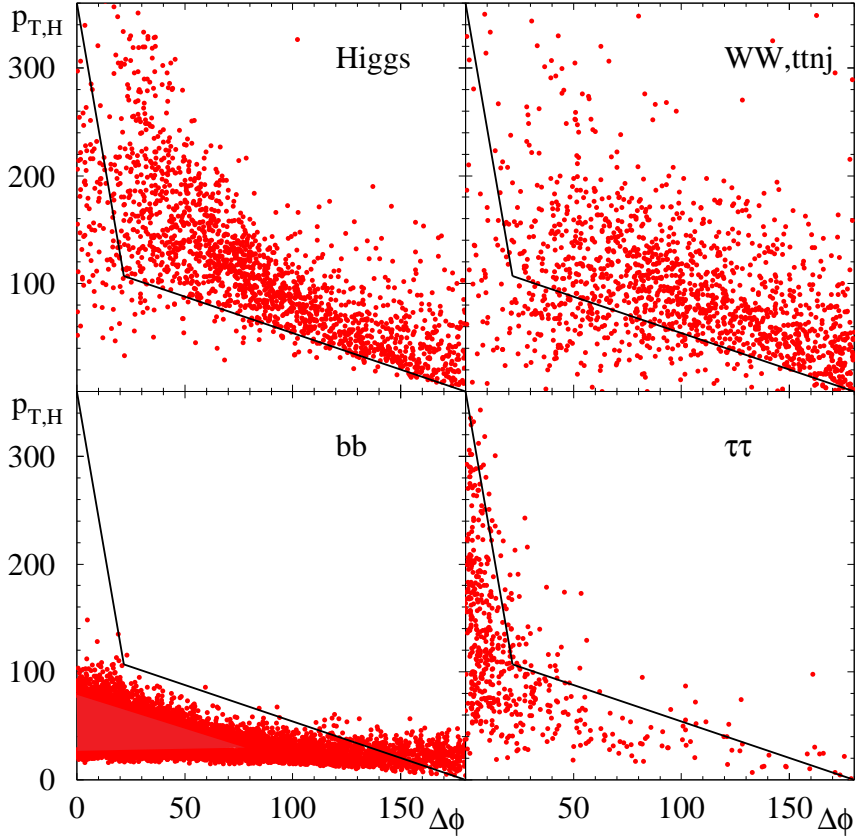} \\
\end{center}
\caption{Distribution of events in the $\Delta\phi(ll,\sla p_T)$ vs. 
$p_{TH}$ plane, where $\Delta\phi(ll,\sla p_T)$ is the azimuthal angle 
between the dilepton momentum and the missing transverse momentum. 
Event numbers correspond to 2000~fb$^{-1}$ and the cuts of 
Eqs.~(\ref{eq:basic}--\ref{eq:MTWW.cut}) and include suppression factors
from a central jet veto on extra parton radiation above $p_T=20$~GeV.
The four panels represent
a $m_H=115$~GeV signal, the combined $W^+W^-$ backgrounds 
from $WWjj$ and $t\bar t+$jets sources, the $bbjj$ background and 
the combined $\tau\tau jj$ backgrounds. Events below and to the left of the 
straight lines are eliminated by the contour cuts of Eq.~(\ref{eq:contour}).
}
\vspace{0.5cm}
\label{fig:contour}
\end{figure}

These correlations are clearly visible in the scatter
plots of Fig.~\ref{fig:contour}, which show the expected distribution of 
events for 2000~fb$^{-1}$ of data in the $\Delta\phi(ll,\sla p_T)$ vs. 
$p_{TH}$ plane. While backgrounds arising from $W^+W^-$ decay look very 
similar to the signal (due to the $m_{ll}$ and $M_T(WW)$ cuts of 
Eqs.~(\ref{eq:mll.phi},\ref{eq:MTWW.cut})) the $bbjj$ and $\tau\tau jj$ 
backgrounds are concentrated at small $\Delta\phi(ll,\sla p_T)$ and/or
small $p_{TH}$ and are effectively eliminated, with little loss for the
signal, by imposing the 'contour cuts'
\begin{equation}
\label{eq:contour}
\Delta\phi(ll,\sla p_T) + 1.5p_{TH} > 180\;,\qquad
12\Delta\phi(ll,\sla p_T) + p_{TH} > 360\;,
\end{equation}
with $\Delta\phi(ll,\sla p_T)$ measured in degrees and $p_{TH}$ measured 
in GeV. The resulting cross sections are listed in the second column 
of Table~I.

\begin{figure}[tb]
\vspace{-0.5cm}
\begin{center}
\includegraphics[width=10.0cm]{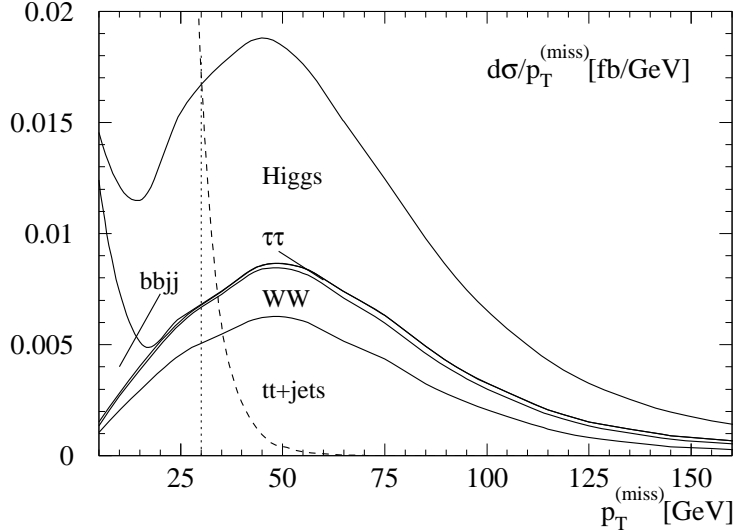} 
\end{center}
\caption{Missing transverse momentum distribution, $d\sigma/d\sla p_T$, after 
the cuts of Eqs.~(\ref{eq:basic}--\ref{eq:contour}) for $l^+l^-\sla p_T$ 
events. The areas between curves represent the contributions from the various 
background classes, as indicated, and the signal. QCD and EW backgrounds
and the $t\bar t$ backgrounds have been combined for clarity. The $l^+l^-jj$ 
background is indicated by the dashed curve. The vertical line represents the 
cut of Eq.~(\ref{eq:cuts.ll}).}
\vspace{0.5cm}
\label{fig:ptmiss}
\end{figure}

It is obvious from Fig.~\ref{fig:contour} that the $bbjj$ background is 
concentrated at low $p_{TH}$: the charm-quarks from $b$-decay will usually 
not pass the lepton isolation cuts of $E_{Tc}<5$~GeV~\cite{RZ_tautau_ll}
unless the parent $b$-quark, and, hence, the decay leptons are soft. This 
leads to a small charged lepton $p_T$ and, simultaneously, to small 
$\sla p_T$ from escaping neutrinos. For large angles 
$\Delta\phi(ll,\sla p_T)$, the missing transverse momentum is dominated by
mismeasurement in the detector, which is small and uncorrelated to the
dilepton direction. The softness of the missing transverse momentum 
distribution for $bbjj$ events, after the contour cuts of 
Eq.~(\ref{eq:contour}), is demonstrated in Fig.~\ref{fig:ptmiss}. 
The remaining $bbjj$ background can be eliminated by a cut
\begin{equation}
\label{eq:emucut}
\sla p_T > 20\;{\rm GeV}\qquad\qquad 
{\rm provided}\; p_{TH}<50\;{\rm GeV}\;.
\end{equation}
Small $\sla p_T$ and small $p_{TH}$ values are largely uncorrelated in the 
signal and the other backgrounds, resulting in a minimal effect of this 
cut (see the third column of Table~I).

The background is dominated by $t\bar{t}j$ production, i.e. events where 
one of the two tagging jets is produced by a $b$-quark from top-decay.
In many cases this jet will be observed in the central region of the
detector, $|\eta_b|<2.5$, where efficient $b$-tagging will be available.
Since the tagging jets of the signal arise from light quarks,
any events with $b$-tagged tagging jets should be eliminated. We assume
a $b$-tagging efficiency of $60\%$ for $b$-jets of $p_{Tb}>20$~GeV in the 
central region ($|\eta_b|<2.5$). This central $b$-veto reduces the $t\bar{t}j$
background by a factor 1.6 and is included in the $t\bar t+$jets cross 
sections of the tables and in the figures. The true $t\bar{t}j$,$t\bar{t}jj$ 
rejection rate will probably be larger, as $b$-quarks of $p_{Tb}<20$~GeV 
can also be recognized via vertex information, although at reduced 
efficiency. Another net gain should arise from vetoing against leptons from 
semileptonic $b$ decays. Our estimates for the backgrounds involving $b$ 
quarks in the final state are conservative, since we do not include
these additional rejection factors.

A further suppression of QCD backgrounds is achieved by a veto on any 
additional central jet activity, i.e. for any jets within the region
defined by Eq.~(\ref{eq:bveto}). In our previous analyses we have estimated
an acceptance of $89\%$ for the WBF signal, $75\%$ for the EW backgrounds and 
$29\%$ for the remaining $t\bar t$ and other QCD 
backgrounds~\cite{RZ_WW,thesis}. These suppression factors are included in 
the fourth column of Table~I and in all figures.

\begin{figure}[tb]
\vspace{-0.5cm}
\begin{center}
\includegraphics[width=10.0cm]{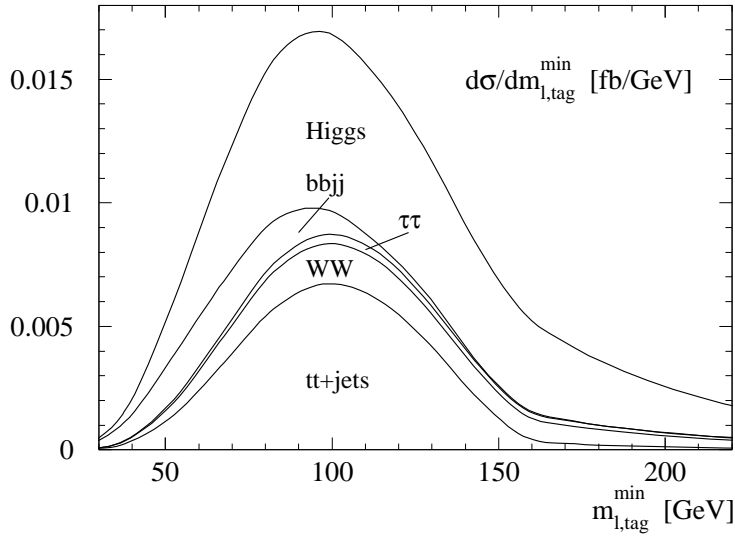} \\
\end{center}
\caption{Distribution of the smallest invariant mass of a 
tagging-jet and a charged-lepton, 
after the cuts of Eqs.~(\ref{eq:basic}--\ref{eq:contour}).
The areas between curves represent the contributions from the various 
background classes, as indicated, and the signal. QCD and EW backgrounds
and the $t\bar t$ backgrounds have been combined for clarity.}
\vspace{0.5cm}
\label{fig:mltag.min}
\end{figure}

In events where one of the tagging jets arises from $t\to bl\nu$ decay, 
the invariant mass of the charged lepton and the $b$-quark jet must be smaller
than the top-quark mass, more precisely $m_{lb}^2 < m_t^2-m_W^2$. Even though
one will not 
know which pair of charged lepton plus tagging jet arises from decay
of a top-quark, the smallest of the four possible combinations, 
$m_{l,tag}^{\rm min}$, has to be below this threshold. This effect is clearly
visible for the $ttj$ background in Fig.~\ref{fig:mltag.min} where we show 
$d\sigma/dm_{l,tag}^{\rm min}$ for various combinations of backgrounds and
the sum of signal plus backgrounds. The region $m_{l,tag}^{\rm min}>155$~GeV
has a very small contribution from top-quark backgrounds and a much improved
S/B ratio could be obtained by a cut. However, neither the significance of 
the signal nor the accuracy of a cross section determination would be improved
by such a cut and therefore we do not consider it any further. The 
$m_{l,tag}^{\rm min}$ distribution contains valuable information for a neural 
net analysis, however, in particular for somewhat larger Higgs boson masses.

\begin{figure}[tb]
\vspace{-0.5cm}
\begin{center}
\includegraphics[width=10.0cm]{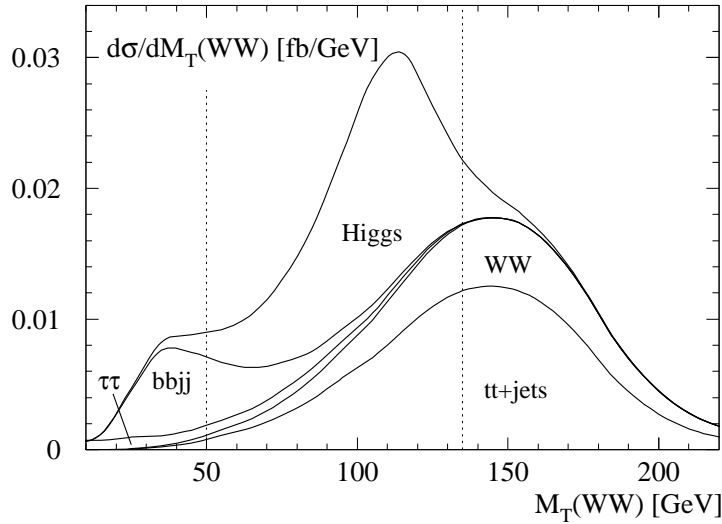} \\
\end{center}
\caption{$WW$ transverse mass distribution, $d\sigma/dM_T(WW)$, after 
the cuts of Eqs.~(\ref{eq:basic}--\ref{eq:contour}) for $e\mu\sla p_T$ events.
The areas between curves represent the contributions from the various 
background classes and the signal, as indicated. The $bbjj$ background is
effectively eliminated by the cut of Eq.~(\ref{eq:emucut}) with minimal effect
on the signal or the other backgrounds.  }
\vspace{0.5cm}
\label{fig:mTWW}
\end{figure}

The cleanest evidence of the signal, and unambiguous information on the 
Higgs boson mass, will be visible in the $WW$ transverse mass distribution,
which is shown in Fig.~\ref{fig:mTWW}. The signal is observed as a clear 
Jacobian peak just below $M_T(WW)=m_H$. One should note that 
the $WW$ transverse mass and the dilepton mass are strongly correlated. 
From Eq.~(\ref{eq:M_T.def}) one immediately finds $M_T(WW)>2m_{ll}$. Thus, 
with $M_T(WW)$ cuts as indicated by the vertical lines in Fig.~\ref{fig:mTWW},
the $m_{ll}<60$~GeV requirement of Eq.~(\ref{eq:mll.phi}) is almost 
automatically fulfilled. The $m_{ll}$ cut does have a strong effect on the
high $M_T(WW)$ range in Fig.~\ref{fig:mTWW}, however.

So far we have only considered the case of two different lepton flavors, i.e.
$H\to WW\to e^\pm\mu^\mp\sla p_T$. Our analysis can be carried over
to $e^+e^-\sla p_T$ and $\mu^+\mu^-\sla p_T$ signatures with only minor 
modifications. Because of the dilepton invariant mass cut, $m_{ll}<60$~GeV,
leptonic $Z$-decays do not pose a large problem. However, low mass 
$l^+l^-$ pairs from $\gamma^*\to ll$ have a large cross section and 
we thus need to impose
a minimum $m_{ll}$ cut, which we choose as 10~GeV. The dominant additional 
background then arises from $lljj$ events where missing transverse momentum 
is generated by detector effects. The contribution of this  
background to the $\sla p_T$ distribution is shown as the dashed line in
Fig.~\ref{fig:ptmiss}. It is tolerable above $\sla p_T=30$~GeV. For the
same flavor case we, thus, impose the additional cuts
\begin{equation}
\label{eq:cuts.ll}
m_{ll}>10~{\rm GeV}\;, \qquad \sla{p}_T > 30~{\rm GeV}\; .
\end{equation}
The resulting signal and background cross sections, including acceptance 
factors for the central jet veto, are listed in the last column of Table~I.

The cross sections obtained in Table~I can be translated into a ``reach''
of LHC experiments for measuring $H\to WW$ in WBF. Quantities of interest
are the minimal luminosity required for a $5\sigma$ observation and the 
statistical accuracy with which the Higgs signal can be determined.
For a realistic estimate we need
to include additional detector effects. As in our previous analyses we assume 
lepton identification efficiencies of 0.95 for each central lepton and
jet reconstruction efficiencies of 0.86 for each of the forward tagging jets,
i.e. the cross sections in Table~I need to be multiplied by an overall
efficiency factor of 0.67. Note that geometric acceptance factors are already
included in the parton level cross sections of Table~I.

In Table~II we separately show the Higgs signal and backgrounds for
the $e\mu \sla p_T$ and the same flavor samples, for Higgs boson masses
between 110 and 140~GeV. The reach of the LHC is given in the last two
columns and combines the information of the two samples. The $5\sigma$ 
discovery luminosity is determined using Poisson statistics for the small
event numbers expected in the initial samples. The last column gives the 
statistical error, $\sqrt{S+B}/S$ which can be expected once both ATLAS 
and CMS have collected 100~fb$^{-1}$ each. The numbers assume a total 
data set of 200~fb$^{-1}$ and thus directly complement the numbers
given in Table~VI of Ref.~\cite{zknr} for the $H\to WW$ decay mode in WBF.


\begin{table*}[t]
\caption{Expected cross sections times detection efficiencies (in fb)
for a variety of Higgs boson masses. The first two columns represent the 
signal S and the sum of all backgrounds, B, as listed in the fourth column
of Table~I for the $e\mu\sla{p}_T$ sample, i.e. the cuts of 
Eqs.~(\ref{eq:basic}--\ref{eq:emucut}) are included. Columns three and four
list analogous results for the $l^+l^-\sla p_T$ case, i.e. within the cuts of
Eqs.~(\ref{eq:basic}--\ref{eq:cuts.ll}). Column five lists the minimal
integrated luminosity required for a $5\sigma$ signal and the last column 
gives the expected statistical error for the determination of 
$B\sigma(qq\to qqH,\;H\to WW)$ with 200~fb$^{-1}$ of data.
}
\label{tab:final}
\begin{tabular}{c|cc|cc|c|c}
&\multicolumn{2}{c|} {$H\to WW\to e^\pm\mu^\mp\sla p_T$} &
 \multicolumn{2}{c|} {$H\to WW\to e^+e^-,\;\mu^+\mu^-+\sla p_T$} & \\ 
 $m_H [{\rm GeV}]$  
       &   $\qquad S\qquad$ &   $\qquad B\qquad$ &   
           $\qquad S\qquad$ &   $\qquad B\qquad$ & 
$5 \sigma\;\int{\cal L}$dt~[fb$^{-1}$]  & accuracy \\
\hline
 110   &  0.30 &  0.43 &  0.21 &  0.39 & 95 &  $16.0\%$   \\ 
 115   &  0.55 &  0.49 &  0.39 &  0.45 & 35 &  $10.3\%$   \\ 
 120   &  0.93 &  0.54 &  0.69 &  0.50 & 15 &  $ 7.1\%$   \\ 
 125   &  1.42 &  0.60 &  1.08 &  0.55 &  8 &  $ 5.4\%$   \\ 
 130   &  2.10 &  0.66 &  1.60 &  0.61 &  4 &  $ 4.3\%$   \\ 
 140   &  3.41 &  0.78 &  2.72 &  0.72 &  2 &  $ 3.2\%$   \\ 
\end{tabular}
\end{table*}

In conclusion, we find that even for a 
SM Higgs boson mass of 115~GeV, the WBF signal
has characteristics which are sufficiently different from the backgrounds
to allow extraction of a $H\to WW\to l^+{l'}^-\sla p_T$ signal with 
$S/B\approx 1/1$ with a total
signal acceptance of about $7\%$. This represents a substantial improvement
over our earlier analysis~\cite{RZ_WW} which arrived at a factor of 1.6
smaller signal rate for the $e\mu\sla p_T$ final state,
with $S/B\approx 0.6$. The main reasons for the improved results are
i) optimization for lower Higgs boson masses, in particular the lowering
of $m_{ll}$ and lepton $p_T$ cuts; ii) making use of correlations between
lepton and missing $p_T$ directions; 
iii) inclusion of same flavor final states, i.e. $e^+e^-\sla p_T$ and
$\mu^+\mu^-\sla p_T$ final states. In addition, the simulation of the 
dominant background, $t\bar tj$ production, has been improved by including 
all off-shell contributions, which raises this background by about $25\%$.

The final result is highly promising. 35~fb$^{-1}$ should be 
sufficient for a $5\sigma$ Higgs boson signal in the $qq\to qqH,\;H\to WW$ 
channel alone, for $m_H=115$~GeV. This required integrated luminosity drops 
very
quickly for somewhat higher mass values. At face value, this weak boson fusion
mode performs better than the inclusive $H\to\gamma\gamma$ search, for which
an integrated luminosity of 45 (70)~fb$^{-1}$ will be required for a 
$5\sigma$ signal in CMS\cite{CMS} (ATLAS\cite{ATLAS}). 
Additional detector effects may lower
these expectations somewhat. On the other hand, the weak boson fusion signal
possesses a wealth of characteristics which can be exploited even better
in a neural net analysis. $H\to WW$ in WBF now appears to be the most 
promising single channel for detection of a SM Higgs boson at the LHC.

\acknowledgements
This research was supported in part by the University of Wisconsin Research
Committee with funds granted by the Wisconsin Alumni Research Foundation and
in part by the U.~S.~Department of Energy under Contract
No.~DE-FG02-95ER40896. Fermilab is operated by URA under DOE contract 
No.~DE-AC02-76CH03000.



\begin{references}

\bibitem{lep2higgs}
P.~Igo-Kemenes, LEP seminar, CERN, Nov.3, 2000;
R.~Barate {\it et al.}  [ALEPH Collaboration],
hep-ex/0011045;
M.~Acciarri {\it et al.}  [L3 Collaboration],
hep-ex/0011043.

\bibitem{one_loop}
H.~E.~Haber and R.~Hempfling, 
Phys.\ Lett.\ {\bf D48}, 4280 (1993);
M.~Carena, J.R.~Espinosa, M.~Quiros, and C.E.M.~Wagner,
Phys.\ Lett.\ {\bf B355}, 209 (1995).

\bibitem{two_loop}
S.~Heinemeyer, W.~Hollik and G.~Weiglein,
Phys.\ Rev.\ {\bf D58}, 091701 (1998);
R.-J.~Zhang, Phys.\ Lett.\ {\bf B447}, 89 (1999).

\bibitem{RZ_WW}
D.~Rainwater and D.~Zeppenfeld,
Phys.\ Rev.\ {\bf D60}, 113004 (1999),  
Erratum-ibid. {\bf D61}, 099901 (2000).

\bibitem{thesis}
D.~Rainwater, PhD thesis, hep-ph/9908378.

\bibitem{zknr}
D.~Zeppenfeld, R.~Kinnunen, A.~Nikitenko and E.~Richter-Was,
Phys.\ Rev.\  {\bf D62}, 013009 (2000).

\bibitem{CMS}
G.~L.~Bayatian {\it et al.}, CMS Technical Proposal,
report CERN/LHCC/94-38 (1994); 
R. Kinnunen and D. Denegri, 
CMS NOTE 1997/057; 
R. Kinnunen and A. Nikitenko, 
CMS TN/97-106; 
R.Kinnunen and D. Denegri, 
hep-ph/9907291.   

\bibitem{ATLAS}
ATLAS Collaboration, ATLAS TDR, 
report CERN/LHCC/99-15 (1999).


\bibitem{kauer}
N.~Kauer and D.~Zeppenfeld, preprint MADPH-00-1205.

\bibitem{RZ_tautau_ll}
T.~Plehn, D.~Rainwater and D.~Zeppenfeld,
Phys.\ Rev.\  {\bf D61}, 093005 (2000).

\bibitem{Cahn}
R.~N.~Cahn, S.~D.~Ellis, R.~Kleiss and W.~J.~Stirling, 
Phys.\ Rev.\ {\bf D35}, 1626 (1987);
V.~Barger, T.~Han, and R.~J.~N.~Phillips, Phys.\ Rev.\ {\bf D37}, 2005 (1988);
R.~Kleiss and W.~J.~Stirling, Phys.\ Lett.\ {\bf 200B}, 193 (1988);
D.~Froideveaux, in {\it Proceedings of the ECFA Large Hadron
Collider Workshop}, Aachen, Germany, 1990, edited by G.~Jarlskog and D.~Rein
(CERN report 90-10, Geneva, Switzerland, 1990), Vol~II, p.~444;
M.~H.~Seymour, {\it ibid}, p.~557;
U.~Baur and E.~W.~N.~Glover, Nucl.\ Phys.\ {\bf B347}, 12 (1990);
Phys.\ Lett.\ {\bf B252}, 683 (1990).

\bibitem{BCHP}
V.~Barger, K.~Cheung, T.~Han, and R.~J.~N.~Phillips,
Phys.\ Rev.\ {\bf D42}, 3052 (1990);
V.~Barger {\it et al.},
Phys.\ Rev.\ {\bf D44}, 1426 (1991);
V.~Barger, 
K.~Cheung, T.~Han, and D.~Zeppenfeld,
Phys.\ Rev.\ {\bf D44}, 2701 (1991);
erratum Phys.\ Rev.\ {\bf D48}, 5444 (1993);
Phys.\ Rev.\ {\bf D48}, 5433 (1993);
V.~Barger {\it et al.},
Phys.\ Rev.\ {\bf D46}, 2028 (1992).

\bibitem{DGOV}
D.~Dicus, J.~F.~Gunion, and R.~Vega,
Phys.\ Lett.\ {\bf B258}, 475 (1991);
D.~Dicus, J.~F.~Gunion, L.~H.~Orr, and R.~Vega,
Nucl. \ Phys.\ {\bf B377}, 31 (1991).

\bibitem{bjgap}
Y.~L.~Dokshitzer, V.~A.~Khoze, and S.~Troian, in {\it
Proceedings of the 6th International Conference on Physics in 
Collisions},
(1986) ed.\ M.~Derrick (World Scientific, 1987) p.365;
J.~D.~Bjorken, Int.\ J.\ Mod.\ Phys.\ {\bf A7}, 4189 (1992);
Phys.\ Rev.\ {\bf D47}, 101 (1993);
V.~Barger, R.~J.~N.~Phillips, and D.~Zeppenfeld,
Phys.\ Lett.\ {\bf B346}, 106 (1995).

\bibitem{RSZ_vnj}
D.~Rainwater, R.~Szalapski, and D.~Zeppenfeld,
Phys.~Rev.~{\bf D54}, 6680 (1996).

\bibitem{DittDrein}
M.~Dittmar and H.~Dreiner, 
Phys.\ Rev.\ {\bf D55}, 167 (1997); 
and [hep-ph/9703401].

\bibitem{tautaumass}
R.~K.~Ellis, I.~Hinchliffe, M.~Soldate and J.~J.~van der Bij,
Nucl.\ Phys.\ {\bf B297}, 221 (1988).

\end{references}
\end{document}